# Quantum "contact" friction: the contribution of kinetic friction coefficient from thermal fluctuations

Rasoul KHEIRI[1*]

[1] Skolkovo Institute of Science and Technology, Moscow, 121205, Russia

* Corresponding author: Rasoul KHERI, E-mail: Rasoul.kheiri@skoltech.ru

**Abstract:** A thermal model of kinetic friction is assigned to a classical loaded particle moving on a fluctuating smooth surface. A sinusoidal wave resembles surface fluctuations with a relaxation time. The Hamiltonian is approximated to the mean energy of the wave describing a system of Harmonic oscillators. The quantization of amplitudes yields in terms of annihilation and creation operators multiplied by a quantum phase. Further, we consider acoustic dispersion relation and evaluate the friction coefficient from the force autocorrelation function. While the sliding particle remains classical describing a nano-particle or a tip with negligible quantum effects like tunneling or delocalization in the wave function, the quantized model of the surface fluctuations results in the temperature dependence of the kinetic friction coefficient. It follows an asymptotic value for higher temperatures and supperslipperiness at low temperatures.

**Keywords:** Kinetic friction coefficient, Fluctuation-dissipation, Quantum friction, Thermal model, phonon

**Nomenclature**

| | |
|---|---|
| $\gamma$ | Kinetic friction coefficient (kg/s) |
| $\Gamma$ | Asymptotic value of $\gamma$ (kg $s^{-1}$ ) |
| $k$ | Wave number ($m^{-1}$) |
| $k_B$ | Boltzmann constant ($m^2$ kg $s^{-2}$ $K^{-1}$) |
| $\rho_s$ | Surface density (kg $m^{-2}$) |
| $S$ | Surface area ($m^2$) |
| $m = \rho_s S$ | Unit mass of the surface (kg) |
| $\omega$ | Frequency ($s^{-1}$) |
| $A$ | Amplitude (m) |
| $\Phi$ , $\phi$ | Phase arguments (unitless) |
| $F_N$ | Normal load (N) |
| $\zeta(t)$ | Random force (N) |
| $F_{fr}$ | Friction force (N) |
| $T$ | Absolute temperature (K) |
| $\varepsilon$ | Mean energy (J) |
| $H$ | Hamiltonian (J) |
| $E_{n_k} = n_k \hbar \omega_k$ | Energy (J) |
| $\in_k = \beta \hbar \omega_k$ | Dimensionless variable |
| Z | Partition function (unitless) |
| $\tau_{ph}$ | Phonon relaxation time (s) |
| $c_s$ | Speed of sound (m/s) |
| $z_p$ | Height of the particle (m) |
| $z_k$ | Height of the surface (m) |
| $\xi$ | Surface deformation (m) |
| $R_p$ | Particle radius (m) |
| $\Theta_D$ | Debye temperature (K) |
| $\hbar$ | Plank constant divided by 2π (Js) |
| *NVE* | Microcanonical ensemble |

## 1 Introduction

Friction at the quantum scale is a broad open problem. For one thing, since the sources of dissipation on solids are a variety including phonons, electrons, adhesion, surface defects in the lattice, chemical bonds, etc., a comprehensive view of quantum friction might be possible by considering only one source of friction in a model and then continuing to sum up with other effects. Some specific quantum phenomena related to friction are quantum tunneling, wave function delocalization, superfluidity, superconductivity, and quantum fields near a surface. Yet, one can also think of using classical models to enhance the quantum description of friction. For example, when a normal load applies to an object, the force on each atom determines by the quantum mechanics of electrons, which is not an easy task to solve in tribology. Then, a synthesis of molecular dynamics (classical) and density-functional (quantum) theory might be a good start [1]. Besides, quantization of a given model from the classical contact mechanics could be significant to introduce quantum “contact” mechanics. In the following, we have a glance at some quantum effects in solid friction and then introduce a quantization for a classical contact model.

The velocity-dependent friction forces in the nanoscale could be classified as adhesion, deformation, and stick-slip forces [32]. In the case of atomic stick-slip where potential barriers are insurmountable for the classical particle, the quantum *tunneling* effect results in permeation of the particle where potential barriers are extending over the kinetic energy of the particle moving on the surface [33-35]. Moreover, in the absence of static friction in incommensurate crystals [18–22], phononic friction for atomically flat dielectrics is temperature-dependent. As a consequence, quantum superslipperiness occurs at low temperatures [23], which is in analogous to the vanishing viscosity in the bulk of fluid that is called *superfluidity* [17]. Furthermore, using Quartz Crystal Microbalance (QCM), J. Krim, et al [24], observed a dramatic reduction in friction at the critical superconducting temperature of the Pd surface when solid nitrogen moved along the surface.[1] Further experiments show the electronic origin of friction in metals in addition to phonons. Remarkably, the non-contact sliding friction coefficient is measured employing a pendulum oscillating on an Nb surface [31]. The temperature is around the critical temperature, and the experiment has done in a high vacuum. As the result, a reduction in the friction coefficient is observed conforming to the electron *superconductivity* of metals.

*Non-contact friction* A fluctuating electromagnetic field near a surface could result in a dissipative part in the van der Waals interaction when an object moving close to the surface [2, 3]. The term "quantum friction" was coined in such pioneering investigations followed by [4–8]. Further studies are inspired mainly by the Casimir effect [9–14] in a dynamic scenario utilizing quantum field theory to address non-contact dynamic models. Yet, other features like exchanging virtual photons with a Doppler shift have also been considered. As a common aspect, non-contact models contain a separation between the moving object, say a particle, and the surface. There are two limitations in this sense. For one thing, a correspondence between microscopic non-contact formalism and classical observations might not be possible. Thus, the quantum roots of classical observations will remain under question. As another constraint, the contribution of the non-contact friction force is experimentally elusive and negligibly much smaller than the contact friction force [15, 16].

The definition of quantum contact friction can be challenging. Nonetheless, the quantization of problems in classical contact mechanics when a normal load is applied to the sliding body on a surface can be regarded as quantum "contact" sliding friction. An example is a classical particle that deforms elastically. In this case, a quantization for a Hertzian problem in tribology could be the onset of the quantum contact friction. One advantage of quantization is that a classical model is acting like a lantern showing up the proper way for quantized solutions. The disadvantage of quantization, on the other hand, is that some effects are inherently quantum, like tunneling and delocalization in the wave function for a single atom,[2] and hence, it is difficult (if doable) to think of a classical counterpart for them!

---

[1] While the observation of a reduction in friction around the critic temperature is impressive, there have been some debates around the "abruptness" of the friction drop [25–30].

Macroscopically, the ”dry friction” between solid surfaces is a force $F_{fr}$ that is linearly dependent on the normal load $F_N$ acting on the bulk ($F_{fr} = \mu F_N$). For a body moving through a viscous fluid with viscosity $\gamma$, on the other hand, the dissipative mechanism is a function of the object’s velocity $\mathbf{v}$, to be linear $F_{fr} = -\gamma \mathbf{v}$ at relatively low speeds. One main difference between dry friction and viscous friction is that in the former, there is static friction ($\mu_s > \mu_k$), which is absent in the latter. In addition, Coulomb’s laws of dry friction emphasize that the friction force is independent of the object’s velocity for ordinary sliding paces as opposed to that of viscous friction.[3] Nevertheless, micro-scale situations are often more detailed and more complicated. For example, regarding a thin film, there might be no constant and net normal force on it. In this case, surface friction can be close to a viscous fluid regime in the absence of any classical fluid [42]. On the atomic scale, while surface asperities are a certain cause of solid friction, the friction coefficient can remain non-zero or even relatively high in the absence of wear, and plowing effects [43]. Equally important, one may find a superlubricity regime for such surfaces as a quantum effect when approaching absolute zero temperature [23]. These two thresholds are of fundamental importance for more investigations.

---

[2] Even in the case of a tip apex’s delocalization [101,102], the Ehrenfest theorem applies.

[3] The autonomy of friction force on surface roughness, however, was bestowed via novel experimental evidence in the last few decades [41].

A first-principle model of lattice vibrations was introduced by Prandtl (known as Prandtl Tomlinson or PT model) [44, 45]. The model contains a point particle moving in a combination of a periodic potential and a harmonic oscillator potential as conservative forces. At the same time, the particle is subjected to a viscous damping force, which is non-conservative. For this reason, the coefficient of periodic force determines the static friction, and the damping force is related to sliding friction. Tabor [46] inspired by Tomlinson's work [47], suggested that in the cases where friction is dependent on speed or temperature, the strain energy of shear distortions is lost in the form of vibrations (phonons).[4] As an extension, thermal fluctuations can be added to the Prandtl-Tomlinson model by a term of random force $\zeta(t)$ and a damping viscous force $F_{fr} = -\gamma \mathbf{v}$ to the equation of motion. The random force satisfies the fluctuation-dissipation theorem, which means $\langle \zeta(t) \rangle = 0$, $\langle \zeta(t)\zeta(t') \rangle = 2\gamma k_B T \delta(t - t')$. Moreover, in the case of metals and semiconductors, both dissipation mechanisms of phonons and electrons excitations might be described by a velocity-dependent frictional force $F_{fr} = -\gamma \mathbf{v}$ by $\gamma = \gamma_{ph} + \gamma_{el}$ where $\gamma_{ph}$, and $\gamma_{el}$ stand for phononic and electronic dissipation factors, respectively. For simplicity, we investigate the effect of thermal fluctuations without additional potential and electronic dissipation. Instead, we introduce a quantization for surface fluctuations utilizing annihilation and creation operators together with a phase operator.

In the current study, thermal fluctuations are supposed on an atomically smooth plane for calculating the kinetic friction coefficient where the friction is assumed to be similar to the linear viscous force [37].[5] On the surface, we consider an elastically deformed particle undergoing a normal load described by a Hertzian force. Moreover, a sinusoidal function approximates the surface fluctuations for evaluating the friction force. Further, the kinetic friction coefficient will be assessed via the fluctuation-dissipation theorem for a near-equilibrium ensemble. As the result, a quantized force brings about a temperature-dependent kinetic friction coefficient. Our work steps are as the following. In Section II, the classical model is elaborated. Next, in Section III, while the tip is still classical, surface fluctuations are described by raising and lowering operators together with a phase operator. Furthermore, the friction force will be quantized from which the kinetic friction coefficient is calculated. In consequence, the temperature-dependent friction coefficient will be evaluated and illustrated. Finally, an appendix expands on the phase operator properties as used in the article, and supplementary material provides more details on the classical derivation.

---

[4] Tabor wrote [46]:*" The basic idea expressed here is not original; it is to be found in a slightly different form in a 1929 paper by Tomlinson [47] where, for some reason, he does not emphasize the important intermediate role of atomic vibrations."*

[5] The classical model was introduced first in [37], and a corresponding mesoscopic model was introduced in [39].

## 2 A Contact model of kinetic friction

Classical contact mechanics is beholden to Heinrich Hertz, who studied elastic deformations published in 1882 on the problem of elastic curved surfaces in contact. Generally speaking, surfaces that deform purely elastically like diamond or carbon coating planes are described by Hertzian force. Possible deviations from Hertzian mechanics are such as plastic deformations, and adhesion in the contact.[6] It is also notable that the exerted normal load can be different from the applied normal load which is assumed to be the same in the Hertzian model.

Experimentally, using atomic force microscopy (AFM) mainly,[7] people have accounted for elastic properties such as Young's modulus of different materials from graphene [50–52] to other substances, including mica [53], and polymers [54]. Such surfaces are suitable as substrates to study elastic properties such as the proportionality of normal load and Hertzian compression [55, 56].[8]

---

[6] A century later [48, 49], a similar theory for adhesive contact was proposed by Johnson, Kendall, and Roberts (JKR-Theory), which reduces to Hertz's theory in the case of zero adhesion. Another model of adhesion is called Derjaguin–Muller–Toporov (DMT) [38], which applies to low adhesion and hard material. An intermediate method is Maugis–Dugdale analytical theory from which JKR and DMT models could be derived as two limits [100].

[7] AFM demonstrated its unique application to study the mechanical and electrical properties of two-dimensional surfaces.

### 2.1 Friction force

Figure 1 is an outline of an ideal flat surface without any other asperities but some sinusoidal vibrations as the only bulges of the surface. In this picture, a particle is being pushed by the normal load $F_z$ while moving on the surface with the velocity $\mathbf{v}$. To begin with, the current height of the surface at point $\boldsymbol{r}$ and time $t$ is determined by the amplitude and phase of the surface phonons at this point, i.e.

$$z_k(\mathbf{r},t) = A_{\mathbf{k}} \cos(\mathbf{k}.\mathbf{r} - \omega_k t + \phi_{\mathbf{k}}) e^{-t/\tau_{ph}}, \tag{1}$$

where $k$ is a given wavenumber ( $k = 2\pi/\lambda$ ), and $\tau_{ph}$ stands for the relaxation time of thermal fluctuations or phonons. In one dimension, the exponential decomposition of Eq. (1) at $t = 0$ is

$$z_k(x,0) = \frac{A_k}{2}[e^{i\Phi_k} + e^{-i\Phi_k}], \qquad \Phi_k = kx + \phi_k. \tag{2}$$

[8] Notably, the graphene monolayer is known as one of the stiffest substrates ever seen [50]. It follows that elastic distortions like wrinkles and similar deformations add friction with fewer atomic layers [57-60]. Since then, molecular dynamics simulations have also investigated the experimentally observed phenomenon [61, 62].

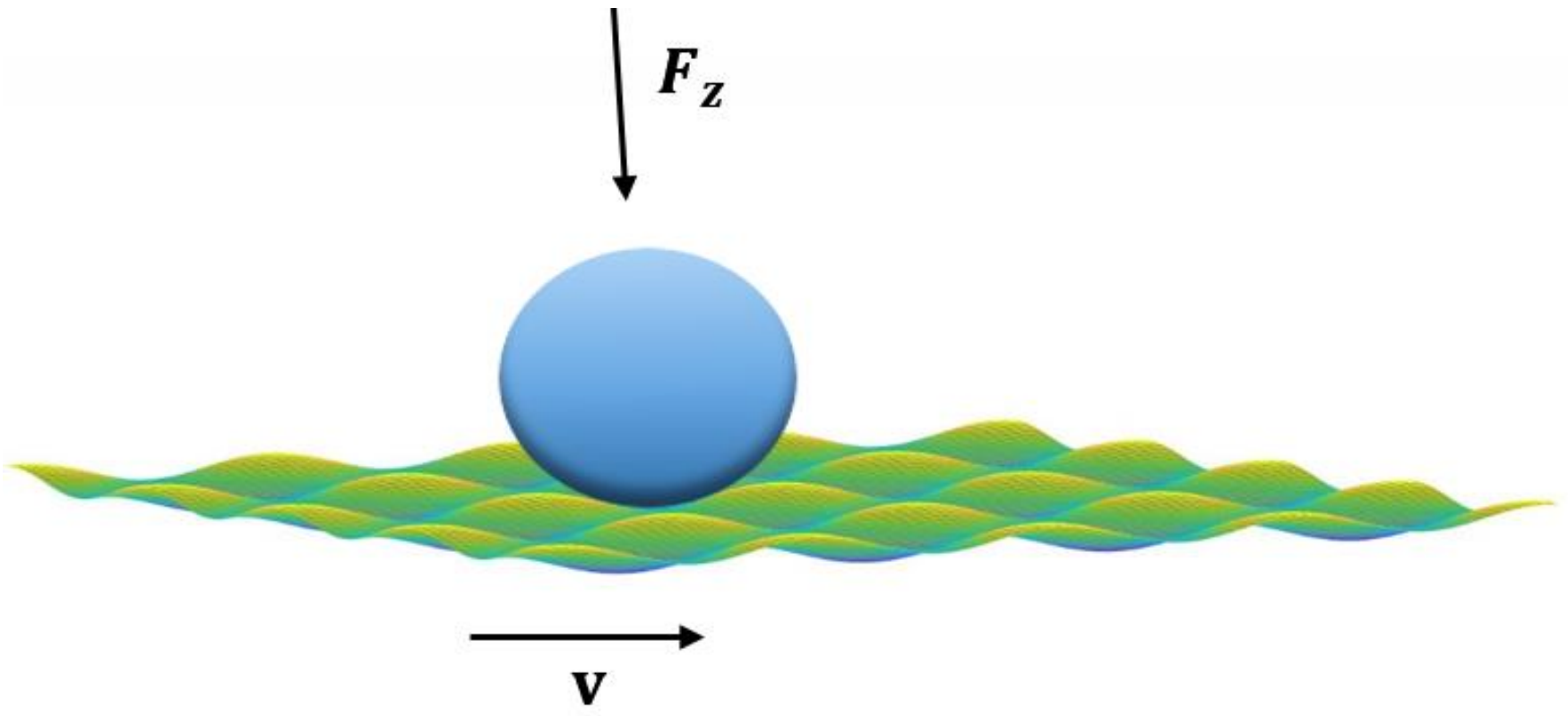

Fig. 1 A spherical tip on a fluctuating surface is squeezed by a normal load $F_z$ while moving with the velocity of $\mathbf{v}$. The fluctuations are described by a cosine wave function and approximated further to the first two terms. The quantization in Sec. 3 is about the surface fluctuations but not the sliding spherical particle. That is, the spherical deformable particle is made up of thousands of atoms with negligible quantum effects like tunneling or delocalization in the wave function.

It turns out that, the first approximation for a non-flat surface comes from small phases, $\Phi_k$, to the extent that:

$$e^{i\Phi_k} \approx 1 + i\Phi_k + \frac{i^2\Phi_k^2}{2}, \tag{3}$$

from which

$$z_k(x,0) \approx A_k[1 - \frac{\Phi_k^2}{2}] = A_k[1 - \frac{1}{2}(kx + \phi_k)^2]. \tag{4}$$

Alternatively, the inclination of the surface makes an angle, say $\alpha$, for a particle on it. Since the particle is pushing down with a normal load, a horizontal resistant force of $F_N \sin\alpha$ applies to the compressed particle with slight deformation of $\xi$ on the surface. Therefore, the height $z_p$ of a squeezed particle in an inclined surface will be shifted by

$$z_p(x) = R_p - \sqrt{R_p^2 - x^2} - \xi,$$

where $R_p$ stands for the radius of the particle. Thus, for small deformations of $\xi << R_p$,

$$z_p(x) \approx \frac{1}{2R_p} x^2 - \xi. \tag{5}$$

Therewith, the quadratic equation of

$$z_k(x) = z_p(x), \qquad \rightarrow \qquad ax^2 + bx + c = 0, \tag{6}$$

gives rise to the solution points $x_1$ $x_2$ for the contact line between the surface and the deformed particle on it. Therefore, having the center of the contact line as $x_m = (x_1 + x_2)/2 = \text{-}b \;\; 2a$, an estimation for the inclination angle is

$$sin\alpha \approx \tan\alpha = \frac{x_m}{R_p} = -\frac{1}{2R_p}\frac{b}{a} \approx -A_k k \phi_k. \tag{7}$$

where we omitted the term including $A_k^2$ in the denominator under the small-amplitude assumption. [9] Consequently, the horizontal force against the particle movement reads

$$F_k = F_N \sin\alpha \cong -(A_k k \phi_k) F_N. \tag{8}$$

This is the kinetic friction, $F_{fr} \equiv F_k$, when a particle traverses such a fluctuating surface undergoing a normal load $F_N$. The normal load is assumed to impose small deformations $\xi$ to the particle, and the surface fluctuations are approximated to the small phases $\Phi_k$ according to Eq. (3).

## 2.2 Fluctuation-dissipation theorem

From the velocity of the classical particle, we can define the thermal frictional coefficient via the fluctuation-dissipation theorem [37, 63]

$$\mathbf{F}_{\text{sol.fr}}(t) = \sum_{\mathbf{k}} \mathbf{F}_{\mathbf{k}}(t) = -\int_0^t \tilde{\gamma}_{\text{sol.fr}}(t-\tau)\, \mathbf{v}(\tau) d\tau + \zeta(t),$$

---

[9] Considering the maximum wave numbers, $k_{\max} = K_D$, according to the Debye temperatures $k_B \Theta_D = \hbar\omega_D = \hbar c_s k_D$, one notes that the expression including $R_k \;\; k$ is not more than unity for a nanoparticle.

where $\mathbf{v}$ stands for the velocity of the particle, and $\tilde{\gamma}_{\text{sol.fr}}(t)$ is the memory function. Accordingly, the microscopic random forces $\zeta(t)$ are related to the macroscopic friction coefficient of the solid friction force $\mathbf{F}_{\text{sol.fr}}$. Namely,

$$\langle \zeta(t^{'}) \cdot \zeta(t) \rangle = 2k_B T \tilde{\gamma}_{\text{sol.fr}}(t-t^{'}), \qquad \langle \zeta(t) \rangle = 0,$$

where $k_B$ and $T$ are Boltzmann constant and temperature, respectively. For non-correlated random forces, $\tilde{\gamma}_{\text{sol.fr}}(t-t^{'}) = 4\gamma_{\text{sol.fr}}\delta(t-t^{'})$, the two-dimensional friction coefficient will be evaluated by

$$\gamma_{sol.fr} = \frac{1}{2k_B T} \int_0^\infty \sum_{\mathbf{k_1},\mathbf{k_2}} \left\langle \mathbf{F_{k_1}(0)}.\mathbf{F_{k_2}(t)} \right\rangle dt.$$

To calculate the time correlation function, averaging takes place to the canonical (Boltzmann-Gibbs) distribution of

$$\exp(-\beta H)/Z, \tag{9}$$

which implies that the system with Hamiltonian $H$ is in thermal equilibrium. Here $Z$ stands for the partition function and $\beta = 1/k_B T$. In addition, assuming isotropic directions for wavenumbers, the friction coefficient reduces to

$$\gamma_{sol.fr.} = \frac{1}{2k_B T} \int_0^\infty \sum_{\mathbf{k}} \left\langle \mathbf{F_k(0)}.\mathbf{F_k(t)} \right\rangle dt, \tag{10}$$

where the wavevector $\mathbf{k}$ in the summation refers to the number of wavenumbers $k$ on a two-dimensional surface.

The thermal friction coefficient assumes to satisfy the fluctuation-dissipation theorem with a zero mean,[10] when a Langevin equation describes the motion of a particle, including a random thermal force [64]. In our case, $F_k$ is approximated by Eq. (8) in each direction. In brief, the thermal frictional force $-\gamma_{(sol.fr.)}.\mathbf{v}$ is acting against the velocity $\mathbf{v}$ of the squeezed particle. It has a different

---

[10] Prandtl himself seemingly considered the first thermal effects in his 1928 paper [44,45]

nature other than the Coulomb friction force, which is athermal and does not depends on particle velocity. The diffusion coefficient, in turn, is obtainable via the corresponding Langevin equation [37].

### 2.3 Mean energy

For a sinusoidal wave in two dimensions

$$z_k(\mathbf{r},t) = A_{\mathbf{k}} \cos(\mathbf{k}.\mathbf{r} - \omega_k t + \phi_{\mathbf{k}})$$

the wave mean energy might be obtained by $\varepsilon_k = \rho_s S \langle (\partial z / \partial t)^2 \rangle$ [65], from which

$$\varepsilon_k = \frac{1}{2} \rho_s S \omega_k^2 A_k^2. \tag{11}$$

Therefore, considering a system of non-interacting lattice excitations, the mean energy obtains form

$$\varepsilon = \frac{1}{2} \rho_s S \sum_{k=1}^{k_D} \omega_k^2 A_k^2. \tag{12}$$

We use the mean energy for the classical averaging. By quantization, we show that the mean energy corresponds to the Hamiltonian of a system of Harmonic oscillators.

#### 2.3.1 Thermal equilibrium assumption

Generally speaking, friction is a non-equilibrium phenomenon. In particular, Eq. (1) describes non-equilibrated phonons whereas we take averages in equilibrium by Eq. (9). To take the average of Eq. (10), we may assume that the surface is in thermal equilibrium with small negligible fluctuations. Such a surface might be constructed in molecular dynamics simulation using a Langevin thermostat. Indeed, utilizing the Langevin thermostat plus NVE integration results in a canonical ensemble with nearly conserved energy [66]. However, the particle itself moves, and external forces are exerted on it. Then, after all, how can we justify an averaging in the thermal equilibrium for the whole system?

For out-of-equilibrium systems, one may imagine the system in a local equilibrium thermalizing with its immediate surrounding. However, it is shown that the local equilibrium assumption can still fail in quantum friction [67].

To put it another way, for addressing the above question, first of all, if the particle finally reaches a constant velocity regime, then it will be in a steady state. For the second, we can imagine thermostats for keeping not only the surface but also the particle at a (nearly) constant temperature. Although such ideal thermostats might not always be realistic, our assumption reduces to the thermal equilibrium average for a particle in a steady state when the whole system is in (nearly) constant temperature.

### 2.4 Classical friction coefficient

Substitution of Eq. (8) in Eq. (10) gives rise to

$$\gamma_{sol.fr.} = \frac{F_N^2}{2k_B T}\int_0^{\infty} dt \sum_{\mathbf{k}} k^2 \left\langle A_k(0)A_k(t)\right\rangle \left\langle \phi_k(0)\phi_k(t)\right\rangle . \tag{13}$$

On the surface of S [68]

$$\sum_{\mathbf{k}} f(k) = \frac{S}{2\pi}\int_0^{k_D} f(k)k\,dk, \tag{14}$$

where $k_D$ is the maximum wavenumber. Moreover, the time dependency is a relaxation for the amplitudes, $A_k(t) = A_k(0)e^{-t/\tau_{ph}}$, and a shift for the phases $\phi_k(t) = \phi_k(0) - \omega_k t$, according to Eq. (1). Therefore,

$$\left\langle A_k(0)A_k(t)\right\rangle = \left\langle A_k^2\right\rangle e^{-t/\tau_{ph}}, \tag{15}$$

and in a symmetric interval

$$\left\langle \phi_k(0)\phi_k(t)\right\rangle = \left\langle \phi_k^2(0)\right\rangle . \tag{16}$$

We take averaging in a thermal equilibrium supposing thermal fluctuations of the surface around a constant temperature. Thereby, an average on the amplitudes $A_k$ makes true in regard to $\mathcal{H}_k$ of the surface fluctuations. Having

$$\varepsilon_k = \frac{1}{2}\rho_s S\omega_k^2 A_k^2, \tag{17}$$

and the definition of

$$\alpha_k^2 = \rho_s S\omega_k^2 / 2k_B T, \tag{18}$$

then,

$$\alpha_k^2 A_k^2 = \beta\varepsilon_k. \tag{19}$$

Furthermore, the mean energy $\varepsilon_k$ is independent of phases. Thereupon, for averaging on phases $\phi_k$

$$\langle\phi_k(0)\phi_k(t)\rangle = \langle\phi_k^2(0)\rangle = \frac{\int_{-\pi}^{\pi}\phi_k^2 d\phi_k}{\int_{-\pi}^{\pi} d\phi_k} = \frac{\pi^2}{3}. \tag{20}$$

It is noted that $\Phi_k = kx + \phi_k$ is small, but $\phi_k \in [-\pi, \pi]$ for which $\langle\phi_k\rangle = 0$. Additionally, in the linear approximation of the acoustic branch, $\omega_k = c_s k$, the phase velocity equals the speed of sound. Eventually, the friction coefficient reads

$$\gamma_{sol.fr.} = \frac{\pi}{12}\frac{F_N^2\tau_{ph}k_D^2}{\rho_s c_s^2}\langle\alpha_k^2 A_k^2\rangle, \tag{21}$$

and

$$\langle\alpha_k^2 A_k^2\rangle = \frac{\int_0^{large}\alpha_k^2 A_k^2 \exp(-\alpha_k^2 A_k^2) d(\alpha_k A_k)}{\int_0^{large}\exp(-\alpha_k^2 A_k^2) d(\alpha_k A_k)} = \frac{1}{2}, \tag{22}$$

where $\alpha_k^2 A_k^2 = \beta\varepsilon_k$ is a dimensionless variable. The upper limit of the integral in Eq. (22) is considered to be large enough for a Gaussian approximation. For one thing, the frequency of phonons in a crystal might be estimated around $(10^{10} - 10^{14})s^{-1}$ [69]. For the second, the red plot in Fig. S1 in the Electronic Supplementary Material shows that a dimensionless value around three is large enough for the upper limit of the integral in Eq. (22). Therefore, we let

$$\Gamma = \frac{\pi}{24} \frac{F_N^2 \tau_{ph} k_D^2}{\rho_s c_s^2}. \tag{23}$$

Here $\Gamma$ refers to the asymptotic value of the kinetic friction coefficient obtained from a Gaussian integration. One can rewrite the asymptotic (saturated) friction coefficient in terms of Debye temperature $\Theta_D$ that could be more tangible than the maximum wavenumbers $k_D$. That is,

$$\Gamma = \left(\frac{k_B}{\hbar}\right)^2 \frac{\pi}{24} \frac{F_N^2 \tau_{ph} \Theta_D^2}{\rho_s c_s^4} \tag{24}$$

with $\hbar c_s k_D = k_B \Theta_D$.

In the next section, we calculate the corresponding quantum friction coefficient utilizing a quantized force. We will show that the friction coefficient is temperature-dependent and superslipperiness occurs at low temperatures.

## 3 Force quantization and friction coefficient

At the atomic level, kinetic energy is transformed into thermal energy [42], and phonons' contribution looks significant in the case of smooth insulators (dielectrics). The simplicity of (non-magnetic) insulators as a substrate to be compared to solid metals, is due to phonons being the only source of transport processes [23, 70]. In this regard, one may assume an ideal surface without any source of friction but thermal phonons. Otherwise, for a more complicated surface, we can refer to the thermal contribution of kinetic friction as one source of dissipation on a given surface.

### 3.1 Quantum amplitudes

For amplitude quantization, it is essential to have surface excitations in correspondence with the classical cosine function of Eq. (1). For this purpose, one can use rising and lowering operators as well as quantum phase operators to build a quantized form of Eq. (1). The phase operator method has firstly been in optics for electromagnetic fields [71, 72]. Here, we suggest a similar theory for phonon excitations on condensed matter.

Assuming surface excitations as

$$\hat{z}(\mathbf{r},t)=\sum_{\mathbf{k}}\sqrt{\frac{\hbar}{m\omega_k}}[\frac{\hat{a}_{\mathbf{k}}e^{i(\mathbf{k}.\mathbf{r}-\omega_k t)}+\hat{a}_{\mathbf{k}}^{\dagger}e^{-i(\mathbf{k}.\mathbf{r}-\omega_k t)}}{2}], \tag{25}$$

we show that such excitations correspond to the classical wave of Eq. (1). For a single excitation, and in the one-dimensional case at $t=0$

$$\hat{z}_k(x,0)=\sqrt{\frac{\hbar}{m\omega_k}}[\hat{a}_k e^{ikx}+\hat{a}_k^{\dagger}e^{-ikx}]. \tag{26}$$

Now, let's check the classical correspondence with $\hat{a}_k^{\dagger}\to a_k^{*}$. Introducing the real $A_k$ such that

$$a_k=R_k e^{i\phi_k},\qquad a_k^{*}=R_k e^{-i\phi_k} \tag{27}$$

then, classically

$$z_k(x,0)=\sqrt{\frac{\hbar}{m\omega_k}}R_k[e^{i(kx+\phi_k)}+e^{-(ikx+\phi_k)}], \tag{28}$$

which is precisely in correspondence with the cosine function of

$$z_k(x,0)=A_k\cos(kx+\phi_k),\qquad A_k=R_k\sqrt{\frac{\hbar}{m\omega_k}}. \tag{29}$$

Thereupon, we rewrite the Eq. (27) for quantum amplitudes. Namely,

$$\hat{R}_k=\hat{a}_k e^{-i\hat{\phi}_k},\qquad \text{or}\qquad \hat{R}_k=\hat{a}_k^{\dagger}e^{i\hat{\phi}_k}, \tag{30}$$

where $\hat{\phi}_k$ is a Hermitian quantum phase operator detailed in Appendix. Because of commutation, one notes that these two alternatives for the operator $\hat{R}_k$ are not equivalent, in contrast with the classical case. Therefore, we should count for both possibilities in Eq. (30), That is to say,

$$\hat{A}_{k,1}=\sqrt{\frac{\hbar}{m\omega_k}}\hat{a}_k e^{-i\hat{\phi}_k},\quad \text{and}\quad \hat{A}_{k,2}=\sqrt{\frac{\hbar}{m\omega_k}}\hat{a}_k^{\dagger}e^{i\hat{\phi}_k}. \tag{31}$$

Hence, the quantization of amplitudes as a vertical displacement of the surface yields in terms of annihilation $\hat{a}_k$ and creation $\hat{a}_k^\dagger$ operators multiplied by a quantum phase $e^{\pm i\hat{\phi}_k}$ in correspondence with the classical model.

### 3.2 Approximated Hamiltonian

Having Eq. (31), one can take the averages of amplitudes in a given state $|n_k\rangle$. For $\hat{A}_{k,1}$

$$\left\langle n_k \,|\, \hat{A}_{k,1}\hat{A}_{k,1} \,|\, n_k \right\rangle = \frac{2\hbar}{m\omega_k}\left\langle n_k \,|\, \hat{a}_k e^{-i\hat{\phi}_k}\, \hat{a}_k e^{-i\hat{\phi}_k} \,|\, n_k \right\rangle, \tag{32}$$

we can use the Hermitian property of the amplitude as

$$\hat{a}_k e^{-i\hat{\phi}_k} = e^{i\hat{\phi}_k}\, \hat{a}_k^\dagger, \tag{33}$$

which reads

$$\left\langle n_k \,|\, \hat{a}_k e^{-i\hat{\phi}_k} e^{i\hat{\phi}_k}\, \hat{a}_k^\dagger \,|\, n_k \right\rangle = \left\langle n_k \,|\, \hat{a}_k \hat{a}_k^\dagger \,|\, n_k \right\rangle = \left\langle n_k \,|\, N+1 \,|\, n_k \right\rangle$$

or

$$\left\langle n_k \,|\, \hat{A}_{k,1}\hat{A}_{k,1} \,|\, n_k \right\rangle = \frac{\hbar}{m\omega_k}(n_k+1).$$

In a similar way for another possible amplitude,

$$\hat{A}_{k,2} = \sqrt{\frac{\hbar}{m\omega_k}}\hat{a}_k^\dagger e^{i\hat{\phi}_k},$$

the average in a given state $|n_k\rangle$ derives from

$$\langle n_k \,|\, \hat{a}_k^\dagger e^{i\hat{\phi}_k}\, \hat{a}_k^\dagger e^{i\hat{\phi}_k} \,|\, n_k \rangle = \langle n_k \,|\, \hat{a}_k^\dagger \hat{a}_k \,|\, n_k \rangle = n_k.$$

To consider both kinds of amplitudes in the energy,

$$\langle n_k \,|\, \hat{A}_{k,1}\hat{A}_{k,1} + \hat{A}_{k,2}\hat{A}_{k,2} \,|\, n_k \rangle = \frac{2\hbar}{m\omega_k} n_k, \tag{34}$$

where we omitted the constant factor on the right-hand side belonging to the vacuum state that is canceled by changing the reference of the energy [73]. Accordingly, the energy of the states reads $E_{n_k} = n_k \hbar \omega_k$. Eventually, having the classical mean energy, $\varepsilon_k = (1/2) m\omega_k^2 A_k^2$, the mean energy of the wave describes the Hamiltonian of a system of independent harmonic oscillators.

$$\hat{H} = \sum_{k=1}^{k_D} \hbar\omega(k)\, \hat{a}_k^\dagger\, \hat{a}_k. \tag{35}$$

### 3.2.1 Time evaluation and the relaxation time

The time-evolution of a quantum operator follows from $\hat{A}_k(t) = \hat{U}^\dagger \hat{A}_k(0) \hat{U}$ where in a time-independent potential, the propagator is simply a shift as $\hat{U} = \exp(-i\hat{H}t/\hbar)$. Therefore, regarding the Hamiltonian of Eq. (35), the time-dependent phase operator follows (see Eq. A3 in Appendix. A.2)

$$\hat{\phi}_k(t) = \hat{\phi}_k - \omega_k t. \tag{36}$$

Furthermore, the amplitudes in Eq. (31) are time-invariant operators. For example,

$$\hat{U}^\dagger \hat{a}_k e^{-i\hat{\phi}_k} \hat{U} = \hat{U}^\dagger \hat{a}_k \hat{U} \hat{U}^\dagger e^{-i\hat{\phi}_k} \hat{U},$$

and

$$\hat{a}_k(t) e^{-i\hat{\phi}_k(t)} = \hat{a}_k\, e^{-i\omega_k t}\, e^{-i(\hat{\phi}_k - \omega_k t)} = \hat{a}_k\, e^{-i\hat{\phi}_k}.$$

Then, for the time-independent Hamiltonian of Eq. (35), phonon amplitudes are time-invariant.

Yet, a quantum excited state cannot remain exited forever [74]. Ergo, the relaxation time is supposable by considering $\omega_k \to \omega_k \pm i(t/\tau_{ph})$ only for the argument of $e^{\pm i\hat{\phi}_k}$ respectively, where $\tau_{ph}$ refers to the relaxation time of phonons. This argument agrees with the response theory, where a phase shift corresponds to the imaginary part of the frequency response function. Consequently,

$$\hat{A}_k(t) = \hat{A}_k(0)\, e^{-t/\tau_{ph}}, \tag{37}$$

for both $\hat{A}_{k,1}$, and $\hat{A}_{k,2}$. Otherwise, for a first principle relaxation, a time-dependent Hamiltonian $\hat{H}(t)$, is required which we have not followed here. On this account, the Hamiltonian at $t = 0$ is approximated by Eq. (35). However, it would be different for $t > 0$ so that the phase $\hat{\phi}_k$ of a phonon's amplitude has an imaginary part for relaxation. Indeed, the existence of an effective relaxation time implies that the thermal fluctuations are not quantum adiabatic. It means that moving the particle on the surface is not faster than surface changes. Namely, the surface is dynamic during an experiment.

### 3.2.2 Quantum time-correlation function

Having the Hamiltonian of Eq. (35), the aim is to derive the force correlation function and evaluate Eq. (10) in the quantized form. Accordingly, in a quantum regime, the corresponding average of an operator $\hat{A}_k$ in its matrix representation can be obtained by taking a trace weighted by a density operator in equilibrium ( $\hat{\rho}_{eq}$ ). That is

$$\langle \hat{A}_k \rangle = \mathrm{Tr}(\hat{\rho}_{eq} \hat{A}_k), \qquad Z_k = \mathrm{Tr}(e^{-\beta \hat{H}_k}).$$

Thereby, $\langle \hat{A}_k \rangle = \sum_{n_k=0}^{\infty} (e^{-\beta E_{n_k}} / Z_k) \langle n_k | \hat{A}_k | n_k \rangle$ , and the time-autocorrelation function reads

$$C_{\hat{A}_k, \hat{A}_k}(t) = \frac{\sum_{n_k=0}^{\infty} e^{-\beta E_{n_k}} \langle n_k | \hat{A}_k(t) \hat{A}_k(0) | n_k \rangle}{\sum_{n_k=0}^{\infty} e^{-\beta E_{n_k}}}. \tag{38}$$

As a consequence, regarding a Hermitian quantum force $\hat{F}_k^{\dagger} = \hat{F}_k$ , the quantized form of Eq. (10) by $|\mathbf{F}_k| \equiv F_k$ will be

$$\gamma_{sol.fr.} = \frac{1}{2k_B T} \int_0^{\infty} \sum_{\mathbf{k}} \left\langle \frac{\hat{F}_k(0)\hat{F}_k(t) + \hat{F}_k(t)\hat{F}_k(0)}{2} \right\rangle dt. \tag{39}$$

In what follows, we are about to quantize the force of Eq. (8), and evaluate the kinetic friction coefficient in a (nearly) thermal equilibrium.

### 3.3 Quantized friction force

For elastic Hertzian deformation $\xi$, the normal load $F_N = \kappa \xi^{3/2}$ remains classical with the assumption of $A_k << \xi$. In other words, we quantize the surface but not the particle moving on it. In this regard, while the spherical particle in Fig. 1 is a tip or a nanoparticle with negligible quantum effect, the sliding takes place on a quantized fluctuating surface. Thereupon, a quantum force corresponding to Eq. (8) is

$$\hat{F}_k = -(\hat{A}_k k \hat{\phi}_k) F_N. \tag{40}$$

Further, regarding Eq. (37),

$$[\hat{A}_k, \hat{A}_k(t)] = 0.$$

As a result, the quantized force commute at different times, and the friction coefficient of Eq. (39) reduces to

$$\gamma = \frac{1}{2k_B T} \int_0^\infty \sum_{\mathbf{k}} \left\langle \hat{\mathbf{F}}_{\mathbf{k}}(\mathbf{0}).\hat{\mathbf{F}}_{\mathbf{k}}(\mathbf{t}) \right\rangle dt \tag{41}$$

Substituting Eq. (40) to Eq. (41) gives rise to the friction coefficient for the quantized force.

$$\gamma = \frac{F_N^2}{2k_B T} \int_0^\infty \sum_{\mathbf{k}} k^2 \left\langle \hat{A}_k \hat{\phi}_k \hat{A}_k(t) \hat{\phi}_k(t) \right\rangle dt, \tag{42}$$

where $\hat{A}_k(t) = \hat{A}_k(0) e^{-t/\tau_{ph}}$, and $\hat{\phi}_k(t) = \hat{\phi}_k - \omega_k t$. Following the discussion in Appendix. A.3 for $\langle \hat{\phi}_k \rangle = 0$,

$$\left\langle n_k \,|\, \hat{A}_k \hat{\phi}_k \hat{A}_k(t) \hat{\phi}_k(t) \,|\, n_k \right\rangle = \frac{\pi^2}{3} \frac{\hbar e^{-t/\tau_{ph}}}{m\omega_k} n_k \left[ 1 - \frac{\Psi(1, n_k + 1)}{\pi^2 / 3} \right]. \tag{43}$$

where $\Psi(\mathrm{m}, \mathrm{x})$ is the polygamma function of order $\mathrm{m}$, that is the $(\mathrm{m}+1)$ th derivative of the logarithm of the gamma function ($\frac{d^{\mathrm{m}+1}}{d\mathrm{x}^{\mathrm{m}+1}} \ln \Gamma(\mathrm{x})$).

In the thermal equilibrium,

$$\frac{\sum_{n_k}^{\infty} e^{-\beta E_{n_k}} \left\langle n_k \mid \hat{A}_k \hat{\phi}_k \hat{A}_k(t) \hat{\phi}_k(t) \mid n_k \right\rangle}{\sum_{n_k}^{\infty} e^{-\beta E_{n_k}}},$$

and the averages on a given state $|n_k\rangle$ will be derived by Eq. (43).

### 3.4 Friction coefficient

To have a close analytic form, we note that the term $\Psi(1, n_k + 1)/(\pi^2/3)$ in Eq. (43) will rapidly vanish. Then, an approximation comes true when one let

$$\left\langle n_k \mid \hat{A}_k \hat{\phi}_k \hat{A}_k(t) \hat{\phi}_k(t) \mid n_k \right\rangle = \frac{\pi^2}{3} \frac{\hbar e^{-t/\tau_{ph}}}{m\omega_k} n_k, \tag{44}$$

from which the thermal average of the amplitudes obtains simply by

$$\left\langle \hat{A}_k \hat{\phi}_k \hat{A}_k(t) \hat{\phi}_k(t) \right\rangle = \frac{\pi^2}{3} \frac{\hbar e^{-t/\tau_{ph}}}{m\omega_k} \frac{\sum_{n_k} e^{-\beta E_{n_k}} n_k}{\sum_{n_k} e^{-\beta E_{n_k}}} = \frac{\pi^2}{3} \frac{\hbar e^{-t/\tau_{ph}}}{m\omega_k} \frac{1}{e^{\beta\hbar\omega} - 1}. \tag{45}$$

Further, substituting Eq. (45) into Eq. (42) gives rise to

$$\gamma = \frac{\pi^2}{3} \frac{F_N^2 \tau_{ph}}{2m} \sum_{\mathbf{k}} \frac{k^2}{\omega_k^2} \frac{\beta\hbar\omega_k}{e^{\beta\hbar\omega_k} - 1}.$$

For acoustic phonons $\omega_k = c_s k$,

$$\gamma = \frac{\pi^2}{3} \frac{F_N^2 \tau_{ph}}{2mc_s^2} \sum_{\mathbf{k}} \frac{\beta\hbar c_s k}{e^{\beta\hbar c_s k} - 1}.$$

On the surface of S [68]

$$\sum_{\mathbf{k}} f(k) = \frac{S}{2\pi} \int_0^{k_D} f(k) k\, dk,$$

thus

$$\gamma = \frac{\pi}{12} \frac{F_N^2 \tau_{ph}}{\rho_s c_s^2} \int_0^{k_D} \frac{\beta\hbar c_s k^2}{e^{\beta\hbar c_s k} - 1} dk.$$

where $\rho_s = m / S$ is the surface density. Considering the Debye temperature

$$k_B \Theta_D = \hbar \omega_D = \hbar c_s k_D,$$

one can define

$$\in_k = \beta \hbar c_s k, \qquad \frac{\Theta_D}{T} = \beta \hbar c_s k_D, \qquad x = \Theta_D / T,$$

whence the approximated friction coefficient simplifies to the second Debye function:

$$\gamma = \frac{\pi}{24} \frac{F_N^2 \tau_{ph} k_D^2}{\rho_s c_s^2} \frac{2}{x^2} \int_0^x \frac{\in_k^2}{e^{\in_k} - 1} d \in_k .$$

Remembering the asymptotic friction coefficient $\Gamma$, Eq. (23) and Eq. (24), in the classical derivations and the Debye functions

$$D_n(x) = \frac{n}{x^n} \int_0^x \frac{\in_k^n}{e^{\in_k} - 1} d \in_k ,$$

the friction coefficient reads

$$\frac{\gamma}{\Gamma} = D_2(x). \tag{46}$$

Equation (46) is an approximation for the friction coefficient with quantized amplitudes based on the analytic approximation of Eq. (45). One sees that the exclusion of the vacuum state avoids having infinities in the friction coefficient.

Nevertheless, taking into account the term $\Psi(1, n_k + 1) / (\pi^2 / 3)$ in Eq. (43), it will change the thermal equilibrium of Eq. (45) to

$$\left\langle \hat{A}_k \hat{\phi}_k \hat{A}_k(t) \hat{\phi}_k(t) \right\rangle = \frac{\pi^2}{3} \frac{\hbar e^{-t/\tau_{ph}}}{m \omega_k} \frac{\sum_{n_k} e^{-\beta E_{n_k}} n_k \left[ 1 - \frac{\Psi(1, n_k + 1)}{\pi^2 / 3} \right]}{\sum_{n_k} e^{-\beta E_{n_k}}}, \tag{47}$$

so that the friction coefficient is modified as

$$\frac{\gamma}{\Gamma}=\frac{2}{x^2}\int_0^x \epsilon_k^2 d\epsilon_k \frac{\sum_{n_k} e^{-\epsilon_k n_k} n_k \left[1-\frac{\Psi(1,n_k+1)}{\pi^2/3}\right]}{\sum_{n_k} e^{-\epsilon_k n_k}}. \tag{48}$$

We may call the right-hand side of Eq. (48) as $Dp_2(x)$ due to changing $D_2(x)$ by quantum phase average in Eq. A6. Accordingly, with this correction

$$\frac{\gamma}{\Gamma}=Dp_2(x). \tag{49}$$

To sum up, Eq. (46) was the first derivation for the friction coefficient by the second Debye function $D_2(x)$. Then, we added the quantum average of the phases (Eq. A6) resulting in Eq. (49).

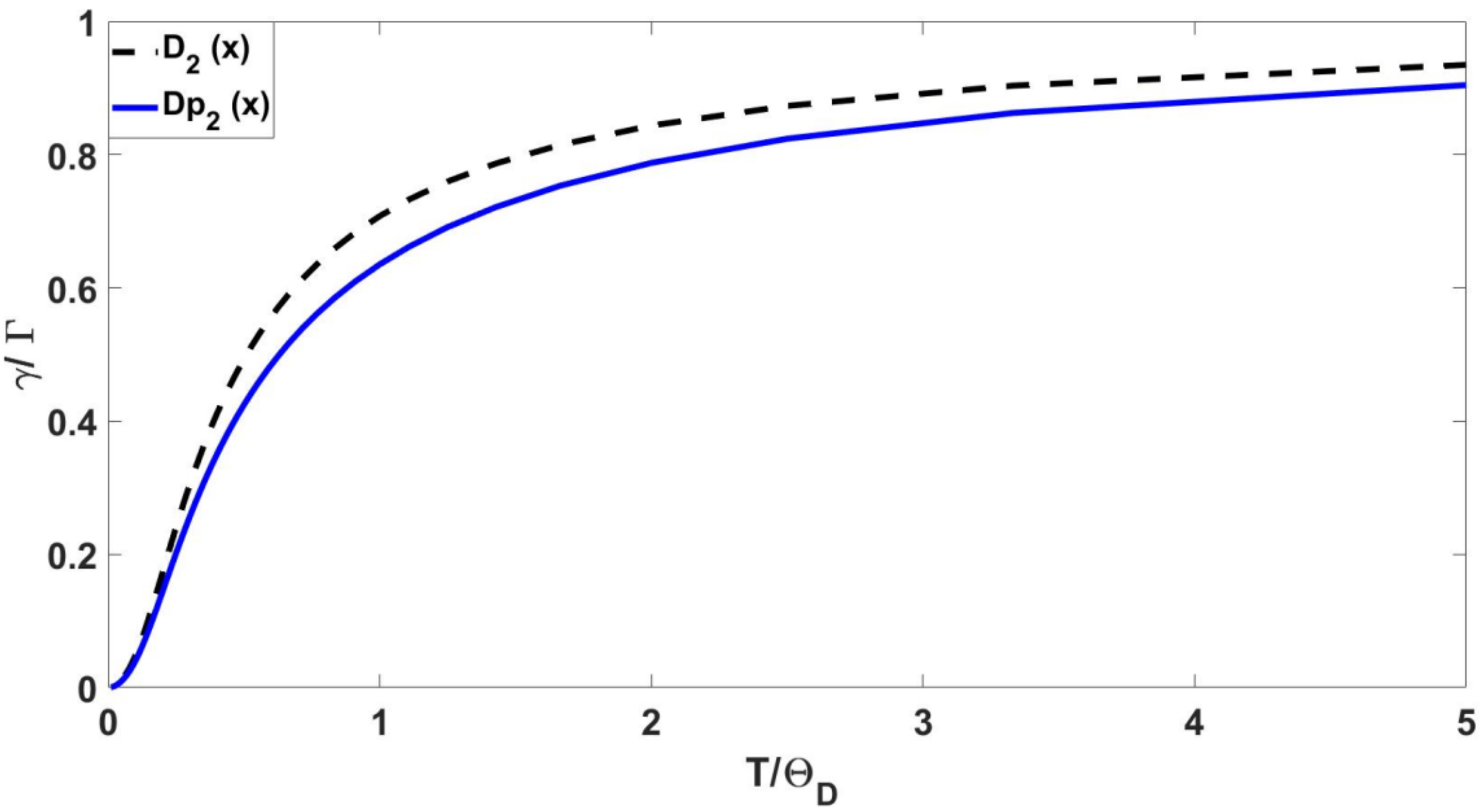

Fig. 2 Kinetic friction coefficient as a function of temperature for the quantized force of Eq. (40). The asymptotic value Γ is identical to the classical case (Eq. (23)). The dashed line refers to the second Debye function $D_2(x)$; an approximation by the quantized amplitudes but classical phases (Eq. (46)). The blue line $Dp_2(x)$ has counted for both quantized amplitudes and phases in Eq. (49).

Figure 2 illustrates the kinetic friction coefficient of Eq. (49) in blue. In addition, the second Debye function, $D_2(x)$, is depicted by a dashed line. The horizontal axis in Fig. 2 is temperature scaled in the Debye temperatures. Thereupon, the friction coefficient vanishes at lower temperatures and reaches the classical asymptotic value for higher temperatures. In the case of $D_2(x)$, the phase average remains identical to the classical phase ($\pi^2/3$). For $Dp_2(x)$, the classical $\pi^2/3$ is replaced by the quantum phase average of Eq. A6. A Comparison between $D_2(x)$ and $Dp_2(x)$ in Fig. 2 shows that the quantum phase average of Eq. A6 decreases the friction coefficient slightly at the middle temperatures.

**4 Conclusion**

In this paper, we followed a model of kinetic friction regarding a thermal-fluctuating surface and a sliding classical spherical particle on it. We considered (acoustic) transversal phonons on the surface where a sinusoidal function with a relaxation time describes the amplitudes. The theory neglects electric charges making the surface more similar to a non-magnetic insulator. In addition, any chemical capacity for particle-surface bonds is not included. ,Moreoverthe surface is assumed to be atomically clean without other surface asperities but thermal fluctuations. Accordingly, the source of dissipation is temperature.

For the sliding particle, we emphasize that it is a classical spherical particle that will be deformed elastically by applying a normal load on it. Therefore, the sliding particle is not a single atom that can tunnel through the surface barriers or it is not even a classical point particle. Henceforth, in the quantization, we quantized the surface fluctuations but not the sliding elastic particle. For this we considered surface amplitudes $A_k$ to be much larger than the radius of the sliding particle $A_k << R_p$ upon which we made an approximation for the friction force in Eq. (8). Further, in quantization, we went a step forward and assumed that $A_k << \xi$. That is, we assumed that the surface amplitudes $A_k$ to be much smaller than the particle deformation $\xi$, from which in the quantization of friction force in Eq. (40), the Hertzman normal load $F_N$ remained classical. In effect, such a particle can be a classical nano-particle, not a quantum particle. Consequently, we quantized the surface fluctuations but not the elastic particle moving on it.

For classical dissipation, a spherical loaded nano-particle is moving on a molecularly smooth surface where the amplitudes of the transversal phonons make the surface inclined. This inclination, in turn, produces a resistance force against the particle's velocity, which is the origin of sliding friction. The friction coefficient is then calculated via the classical fluctuation-dissipation theorem for near-equilibrium linear viscous friction. The averaging of the energy is approximated by the mean energy of the waves corresponding to a system of Harmonic oscillators. Moreover, the resistance force in Eq. (8) is approximated upon small phases, small amplitudes, and lattice wavenumbers multiplied by the normal load exerted on the particle. Thereupon, the classical kinetic friction coefficient was derived, and its asymptotic value in Eq. (23) calculated from a Gaussian approximation.

For a quantum theory, we quantized the friction force through phonons' quantum amplitudes and phases. At the same time, the Hertzian normal load remained classical due to the classical deformations of the sliding particle. That is, by assuming $A_k << \xi$ , we quantized the surface fluctuations but not the sliding particle moving on it. Accordingly, the quantum amplitudes were built by the raising and lowering operators multiplied by a quantum phase operator, $\exp(\pm i\hat{\phi})$. For the quantization of the phases, we postulated $\langle \hat{\phi}_k \rangle = 0$ from a statistical point of view for the wavenumbers $k$ . It is argued that, even if the quantum mechanical phase can be questionable in the ground state, the quantum statistical phase is sensible with $\langle \hat{\phi}_k \rangle = 0$ in a symmetric interval.

In results, we took an approximation for the kinetic friction coefficient by the second Debye function from which the asymptotic value of the friction coefficient is consistent with the classical theory. Next, we added a modification by quantum phase averaging. The resulting friction coefficient is slightly lower than the second Debye function, but with the same asymptotic value, referring to not all phases allowed in the quantization. For $T \to 0$, the model brings about a vanishing friction coefficient, referring to suppersliperiness at low temperatures.

Debye temperature is a finite temperature typically around $(10^2 - 10^3)^{\circ} K$ "above which all modes begin to be excited" [68]. Thus, one may expect that approaching the asymptotic value takes place faster than occurs in Fig. 2, and the friction coefficient would have a higher value around the Debye temperature. Moreover, one notes that the harmonic assumption might not hold when approaching the melting point. Furthermore, there could be a question that why the friction coefficient is reduced by decreasing temperature while it can't be seen in the experimental investigations with AFM. Indeed, thermal activation makes the opposite result of decreasing friction by increasing temperature.

Firstly, we point out that the model as presented in Fig. 2 is an approximation rather than an exact solution. For the second, our model is restricted to only one contribution in the thermal kinetic friction. The other important contribution is referred to thermal activation. In our mind, if the current model will be added to a thermal activation theory, then the result might be more interesting! That is the matter of future studies.

Nevertheless, the advantage of the surface fluctuations' quantization is that the kinetic friction coefficient is derived as a function of temperature with a simple approximation as the second Debye function. The classical derivation, on the other hand, does not provide such a simple one-variable function for the kinetic friction coefficient.

**4.1 Other possibilities for future studies**

we suggest computer simulations of classical molecular dynamics to understand better and confirm the present theory. Having other parameters in the asymptotic value of Eq. (24) for a particular surface, the objective of the simulations is to investigate the classical kinetic friction coefficient as a function of normal load $\Gamma \propto F_N^2$. For one thing, an agreement between the theoretically estimated friction coefficient (a quadratic curve) and simulated points will support the present model.

Furthermore, computer simulations provide more complicated scenarios like rolling friction or adhesion, which can be studied in turn [75, 76].

Last but not least, we end up with a short list of literature that might be considered to enhance the subject of dissipation and quantum friction: The nonlinear dynamics of sliding friction [99], quantum tunneling, and thermal creep [35, 97, 98], Quantum adiabaticity [82], High-temperature superfluidity and bilayer graphene [77-80], Plastic deformations, and serrated deformations at low temperatures [81, 83-85], quantized friction [86], quantum size effect and its connection with adhesion [87, 88], quantum description of fractures in solids [89], quantum random force [90], thermal quantum field theory and thermal states [91, 92].

## Appendix: Quantum phase operator

A quantization for the sinusoidal wave gives rise to the problem of the phase operator. The application has already been in optic regarding the quantization of an electrical plain field [72, 93]. The paradoxical difficulty of the problem is about the ground state $|n=0\rangle$ in quantum mechanics. However, for quantum field states $|n_k\rangle$, the ground state is already excluded in the Hamiltonian by changing the reference to avoid infinities. In this regard, utilizing the quantum phase operator in quantum statistics could be plausible. For literature, one may refer to the matrix representation [94] , a critical review paper [71] , some physics and history of the problem [95] , and a review book [96] .

### A.1 Matrix representation

A quantized phase operator, $\hat{\phi}_k$, in terms of raising and lowering operators will be

$$\exp(i\hat{\phi}_k) = \hat{a}_k \hat{N}_k^{-1/2} = \frac{\hat{a}_k}{\sqrt{\hat{a}_k^\dagger \hat{a}_k}}.$$

One immediate question is about the effect of the phase operator on the states $|n_k\rangle$. Namely,

$$\exp(i\hat{\phi}_k)|n_k\rangle = ?$$

Having $e^{i\hat{\phi}_k}\hat{N}_k^{1/2}=\hat{a}_k$,

$$\langle n_k|e^{i\hat{\phi}_k}\ \hat{N}_k^{1/2}|m_k\rangle=\sqrt{m_k}\langle n_k|m_k-1\rangle,$$

and

$$\langle n_k|e^{i\hat{\phi}_k}\ |m_k\rangle=\langle n_k|m_k-1\rangle=\delta_{n_k,m_k-1}.$$

Therefore, the phase operator is unitary except for the ground state $n_k=0$. Hence, one can write

$$\exp(i\hat{\phi}_k)=\sum_{n_k}|n_k\rangle\langle n_k+1|. \tag{A1}$$

Additionally,

$$[e^{i\hat{\phi}_k},\hat{N}_k]=\sum_{n_k}|n_k\rangle\langle n_k+1|\hat{N}_k-\sum_{n_k}\hat{N}_k|n_k\rangle\langle n_k+1|\ =\sum_{n_k}|n_k\rangle\langle n_k+1|(n_k+1)-\sum_{n_k}(n_k)|n_k\rangle\langle n_k+1|,$$

That is

$$[e^{i\hat{\phi}_k},\hat{N}_k]=\sum_{n_k}|n_k\rangle\langle n_k+1|=e^{i\hat{\phi}_k}.$$

In this representation, we can easily find the dagger operator, as

$$\exp(-i\hat{\phi}_k)=\sum_{n_k}|n_k+1\rangle\langle n_k|,$$

and further,

$$e^{i\hat{\phi}_k}|n_k+1\rangle=|n_k\rangle\qquad\rightarrow\qquad e^{i\hat{\phi}_k}|n_k\rangle=|n_k-1\rangle.$$

To sum up

$$\begin{cases}\exp(i\hat{\phi}_k)|n_k\rangle=|n_k-1\rangle,\\ \exp(-i\hat{\phi}_k)|n_k\rangle=|n_k+1\rangle.\end{cases} \tag{A2}$$

**A.2 Time evolution**

Using the commutator relation

$$[\exp(i\hat{\phi}),N_k]=\exp(i\hat{\phi}_k)\qquad\rightarrow\qquad[N_k,\hat{\phi}_k]=i,$$

and remembering the Hamiltonian, Eq. (35),

$$\hat{H} = \hbar \sum_j \omega_j \hat{a}_j^\dagger \hat{a}_j = \hbar \sum_j \omega_j \hat{N}_j,$$

one can evaluate the time evolution of the phase operator,

$$\hat{\phi}_k(t) = \exp(\frac{i\hat{H}t}{\hbar})\hat{\phi}_k \exp(-\frac{i\hat{H}t}{\hbar}).$$

Using Baker-Hausdorff relation

$$\hat{\phi}_k(t) = \hat{\phi}_k + it\left[\sum_j \omega_j \hat{N}_j, \hat{\phi}_k\right] = \hat{\phi}_k - \omega_j t \delta_{kj}.$$

As a result

$$\hat{\phi}_k(t) = \hat{\phi}_k(0) - \omega_k t, \tag{A3}$$

which is in agreement with the classical $b_k(t) = b_k(0) - ct$ for $\phi_k = kb_k$, where $c = \omega / k$ is the phase velocity.

**A.3 Derivation of Eq. (43)**

The classical phase interval is $\phi_k \in [-\pi, \pi]$, from which $\langle \phi_k \rangle = 0$, and Eq. (20) yielded. Namely,

$$\left\langle \phi_k(0)\phi_k(t) \right\rangle = \left\langle \phi_k^2(0) \right\rangle = \frac{\int_{-\pi}^{\pi} \phi_k^2 \, d\phi_k}{\int_{-\pi}^{\pi} d\phi_k} = \frac{\pi^2}{3}.$$

In the quantum realm, for the commutator $[\hat{N}, \hat{\phi}] = i$, the phase element of $\langle n | \hat{\phi} | n \rangle$ can be undefined or related to a reference in general. However, in the interval of $[-\pi, \pi]$, and for a statistical $k$ number, that is regarding $\phi_k, | n_k \rangle$, but not a given $\phi, | n \rangle$, the reference can be statistical too. For this we postulate $\langle \hat{\phi}_k \rangle = 0$, corresponding to the classical phase interval $k$ for a statistical wave number, $\phi_k \in [-\pi, \pi]$. In this regard, we need to find

$$\langle n_k | \hat{A}_k \hat{\phi}_k \hat{A}_k \hat{\phi}_k | n_k \rangle,$$

For each amplitude

$$\hat{A}_{k,1} = \sqrt{\frac{\hbar}{2m\omega_k}}\hat{a}_k e^{-i\hat{\phi}_k}, \quad \text{and} \quad \hat{A}_{k,2} = \sqrt{\frac{\hbar}{2m\omega_k}}\hat{a}_k^{\dagger} e^{i\hat{\phi}_k}.$$

Let's consider $\hat{A}_{k,1}$, then

$$\langle n_k \mid \hat{A}_{k,1}\hat{\phi}_k \hat{A}_{k,1}\hat{\phi}_k \mid n_k \rangle = \frac{\hbar}{2m\omega_k}\langle n_k \mid \hat{a}_k e^{-i\hat{\phi}_k}\hat{\phi}_k \hat{a}_k e^{-i\hat{\phi}_k}\hat{\phi}_k \mid n_k \rangle,$$

and, $\hat{a}_k = e^{i\hat{\phi}_k}\hat{N}^{1/2}$. Subsequently,

$$\langle n_k \mid \hat{A}_{k,1}\hat{\phi}_k \hat{A}_{k,1}\hat{\phi}_k \mid n_k \rangle = \frac{\hbar}{2m\omega_k}\langle n_k \mid e^{i\hat{\phi}_k}\hat{N}_k^{1/2} e^{-i\hat{\phi}_k}\hat{\phi}_k e^{i\hat{\phi}_k}\hat{N}_k^{1/2} e^{-i\hat{\phi}_k}\hat{\phi}_k \mid n_k \rangle$$

$$= \frac{\hbar}{2m\omega_k}\langle n_k \mid e^{i\hat{\phi}_k}\hat{N}_k^{1/2}\hat{\phi}_k \hat{N}_k^{1/2} e^{-i\hat{\phi}_k}\hat{\phi}_k \mid n_k \rangle$$

from the zero commutators of phase with its exponential function. In the next step, we use

$$[\hat{N}_k, \hat{\phi}_k] = i, \quad \rightarrow \quad [\hat{N}_k^{1/2}, \hat{\phi}_k] = \frac{i}{2}\hat{N}_k^{-1/2},$$

or

$$\hat{N}_k^{1/2}\hat{\phi}_k - \frac{i}{2}\hat{N}_k^{-1/2} = \hat{\phi}_k \hat{N}_k^{1/2}$$

from which

$$\langle n_k \mid e^{i\hat{\phi}_k}\hat{N}_k^{1/2}\hat{\phi}_k \hat{N}_k^{1/2} e^{-i\hat{\phi}_k}\hat{\phi}_k \mid n_k \rangle = \langle n_k \mid e^{i\hat{\phi}_k}\hat{N}_k^{1/2}\left(\hat{N}_k^{1/2}\hat{\phi}_k - \frac{i}{2}\hat{N}_k^{-1/2}\right) e^{-i\hat{\phi}_k}\hat{\phi}_k \mid n_k \rangle$$

$$= \langle n_k \mid e^{i\hat{\phi}_k}\hat{N}_k^{1/2}\hat{N}_k^{1/2}\hat{\phi}_k e^{-i\hat{\phi}_k}\hat{\phi}_k \mid n_k \rangle - \frac{i}{2}\langle n_k \mid e^{i\hat{\phi}_k}\hat{N}_k^{1/2}\hat{N}_k^{-1/2} e^{-i\hat{\phi}_k}\hat{\phi}_k \mid n_k \rangle$$

$$= \langle n_k \mid e^{i\hat{\phi}_k}\hat{N}_k^{1/2}\hat{N}_k^{1/2} e^{-i\hat{\phi}_k}\hat{\phi}_k \hat{\phi}_k \mid n_k \rangle - \frac{i}{2}\langle n_k \mid \hat{\phi}_k \mid n_k \rangle = \langle n_k \mid e^{i\hat{\phi}_k}\hat{N}_k^{1/2}\hat{N}_k^{1/2} e^{-i\hat{\phi}_k}\hat{\phi}_k \hat{\phi}_k \mid n_k \rangle,$$

for $\langle \hat{\phi}_k \rangle = 0$. In addition, we note that,

$$\hat{A}_{k,1}\hat{A}_{k,1} = \frac{\hbar}{2m\omega_k}\hat{a}_k e^{-i\hat{\phi}_k}\hat{a}_k e^{-i\hat{\phi}_k} = e^{i\hat{\phi}_k}\hat{N}_k^{1/2} e^{-i\hat{\phi}_k} e^{i\hat{\phi}_k}\hat{N}_k^{1/2} e^{-i\hat{\phi}_k} = e^{i\hat{\phi}_k}\hat{N}_k^{1/2}\hat{N}_k^{1/2} e^{-i\hat{\phi}_k}.$$

Therefore,

$$\langle n_k \mid \hat{A}_{k,1}\hat{\phi}_k \hat{A}_{k,1}\hat{\phi}_k \mid n_k\rangle = \langle n_k \mid \hat{A}_{k,1}\hat{A}_{k,1}\hat{\phi}_k\hat{\phi}_k \mid n_k\rangle.$$

One can follow a similar argument for $\hat{A}_{k,2}$, and reaches the same result. Thereupon,

$$\langle n_k \mid \hat{A}_k\hat{\phi}_k \hat{A}_k\hat{\phi}_k \mid n_k\rangle = \langle n_k \mid \hat{A}_k\hat{A}_k\hat{\phi}_k\hat{\phi}_k \mid n_k\rangle$$

with $\langle\hat{\phi}_k\rangle = 0$, for both $\hat{A}_{k,1}$ and $\hat{A}_{k,2}$. For the next step,

$$\left\langle n_k \mid \hat{A}_k\hat{A}_k(t)\hat{\phi}_k\hat{\phi}_k(t) \mid n_k\right\rangle = \sum_{n'}\left\langle n_k \mid \hat{A}_k\hat{A}_k(t) \mid n'_k\right\rangle\left\langle n'_k \mid \hat{\phi}_k\hat{\phi}_k(t) \mid n_k\right\rangle$$

$$\left\langle n_k \mid \hat{A}_k\hat{A}_k(t) \mid n'_k\right\rangle = \delta_{nn'}\left\langle n_k \mid \hat{A}_k\hat{A}_k(t) \mid n_k\right\rangle,$$

$$\left\langle n'_k \mid \hat{\phi}_k\hat{\phi}_k(t) \mid n_k\right\rangle = \left\langle n'_k \mid \hat{\phi}_k\hat{\phi}_k \mid n_k\right\rangle.$$

Consequently,

$$\left\langle n_k \mid \hat{A}_k\hat{\phi}_k\hat{A}_k(t)\hat{\phi}_k(t) \mid n_k\right\rangle = \left\langle n_k \mid \hat{A}_k\hat{A}_k(t) \mid n_k\right\rangle\left\langle n_k \mid \hat{\phi}_k\hat{\phi}_k \mid n_k\right\rangle.$$

In the general case of quantum phases [71, 96]

$$\langle n_k \mid \hat{\phi}_k^2 \mid n_k\rangle - \langle n_k \mid \hat{\phi}_k \mid n_k\rangle^2 = \frac{\pi^2}{6} + \sum_{l}^{n_k}\frac{1}{l^2} = \frac{\pi^2}{3} - \sum_{l=n_k+1}^{\infty}\frac{1}{l^2},$$

which for $\langle\hat{\phi}_k\rangle = 0$ reads

$$\langle n_k \mid \hat{\phi}_k^2 \mid n_k\rangle = \frac{\pi^2}{6} + \sum_{l}^{n_k}\frac{1}{l^2} = \frac{\pi^2}{3} - \sum_{l=n_k+1}^{\infty}\frac{1}{l^2},$$

Furthermore,

$$\sum_{l=1}^{n_k}\frac{1}{l^2} = \frac{\pi^2}{6} - \Psi(1, n_k + 1),$$

where $\Psi(\mathrm{m},\mathrm{x})$ is the polygamma function of order $\mathrm{m}$, that is the $(\mathrm{m}+1)$ th derivative of the logarithm of the gamma function ($\frac{d^{\mathrm{m}+1}}{d\mathrm{x}^{\mathrm{m}+1}}\ln\Gamma(\mathrm{x})$). As a result,

$$\langle n_k \,|\, \hat{\phi}_k^2 \,|\, n_k \rangle = \frac{\pi^2}{3} - \Psi(1, n_k + 1). \tag{A4}$$

In the classical limit

$$\lim_{n_k \to \infty} \left\langle n_k \left| \hat{\phi}_k(0)\hat{\phi}_k(t) \right| n_k \right\rangle = \frac{\pi^2}{3}. \tag{A5}$$

which corresponds to Eq. (20). Rewriting Eq. (A4), we have

$$\left\langle n_k \,|\, \hat{\phi}_k(0)\hat{\phi}_k(t) \,|\, n_k \right\rangle = \frac{\pi^2}{3}\left[1 - \frac{\Psi(1, n_k + 1)}{\pi^2/3}\right]. \tag{A6}$$

Finally,

$$\left\langle n_k \,|\, \hat{A}_k \hat{\phi}_k \hat{A}_k(t)\hat{\phi}_k(t) \,|\, n_k \right\rangle = \frac{\pi^2}{3}\frac{\hbar e^{-t/\tau_{ph}}}{m\omega_k} n_k \left[1 - \frac{\Psi(1, n_k + 1)}{\pi^2/3}\right]. \tag{A7}$$

**Acknowledgments**

The author acknowledges Prof. Nikolai Brilliantov for suggesting the problem and thanks his lab members for their discussions. I am also thankful to Prof. Anatoly Dymarsky for our correspondence.

**Declaration of competing interest**

The author has no competing interests to declare that are relevant to the content of this article.

**Author biography**

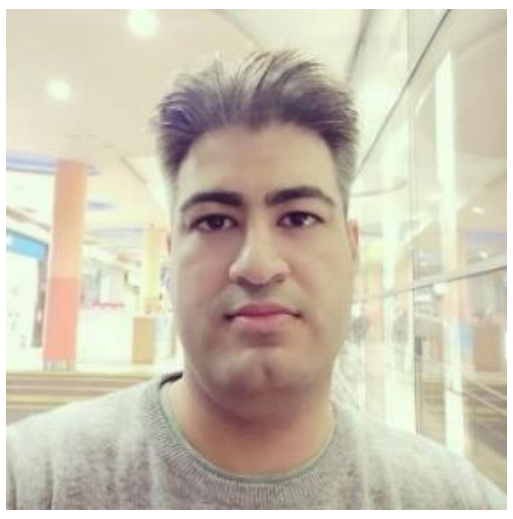

**Rasoul Kheiri** PhD student, obtained his bachelor's degree in physics at IAU, Iran with a first rank award in 2010. He then moved to IUT, Iran for a master's degree in condensed matter physics accomplished in 2015. Afterward, he won a scholarship from Skoltech, Russia in 2018 where he is currently working at. His research interest is quantum physics with two publications so far. In particular, he is interested in the correspondence between quantum and classical systems, fluctuation-dissipation theorem, statistical mechanics, and condensed matter physics. His current project is molecular modeling in nano-tribology supervised by Prof. Nikolai Brilliantov.

**Graphical abstract**

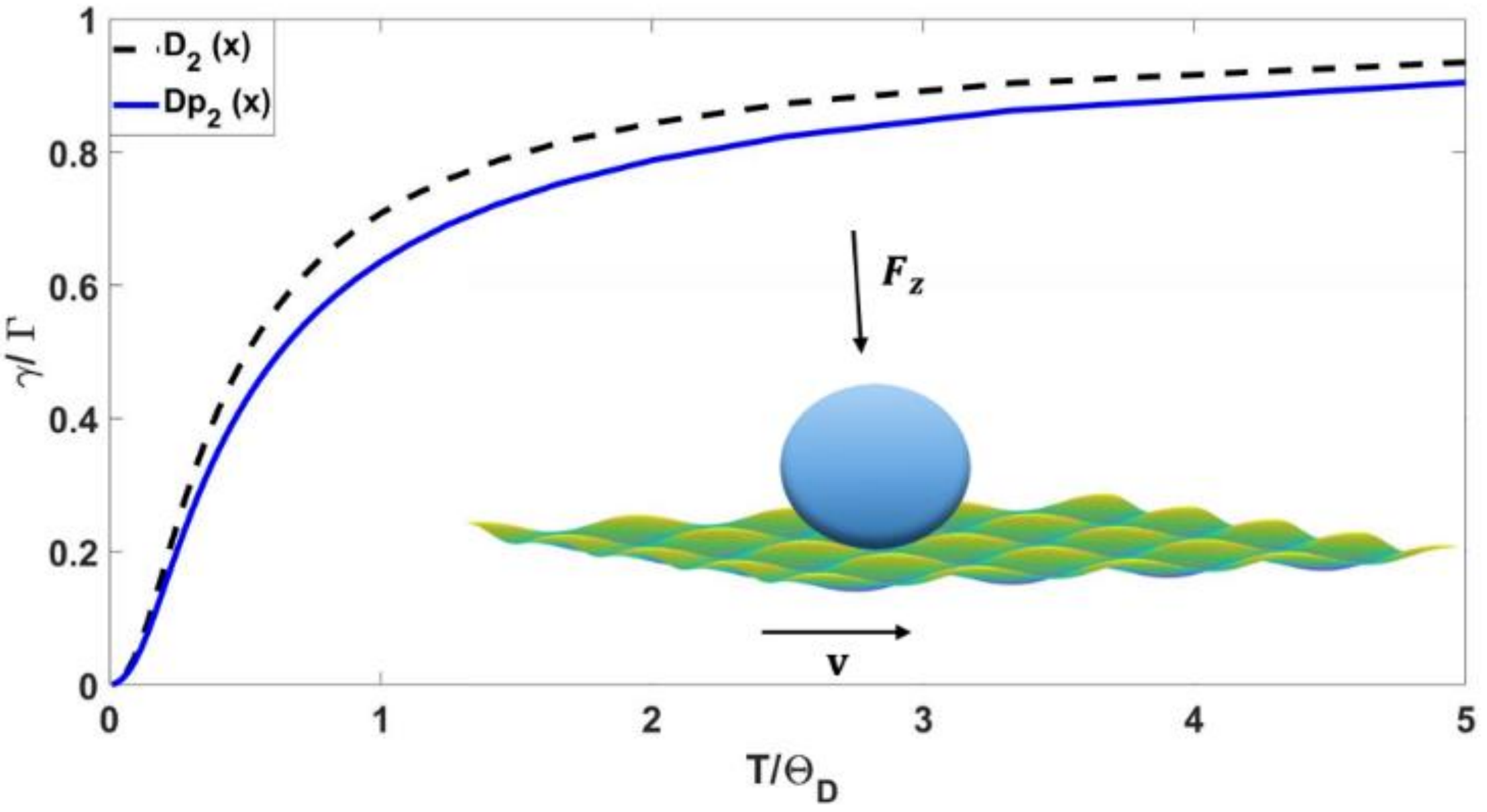

**Electronic supplementary material**

A comparison between classical and quantum derivations of the kinetic friction coefficients could be possible by considering not only the saturated but also unsaturated cases of the classical friction coefficient in the integral of Eq.(22).

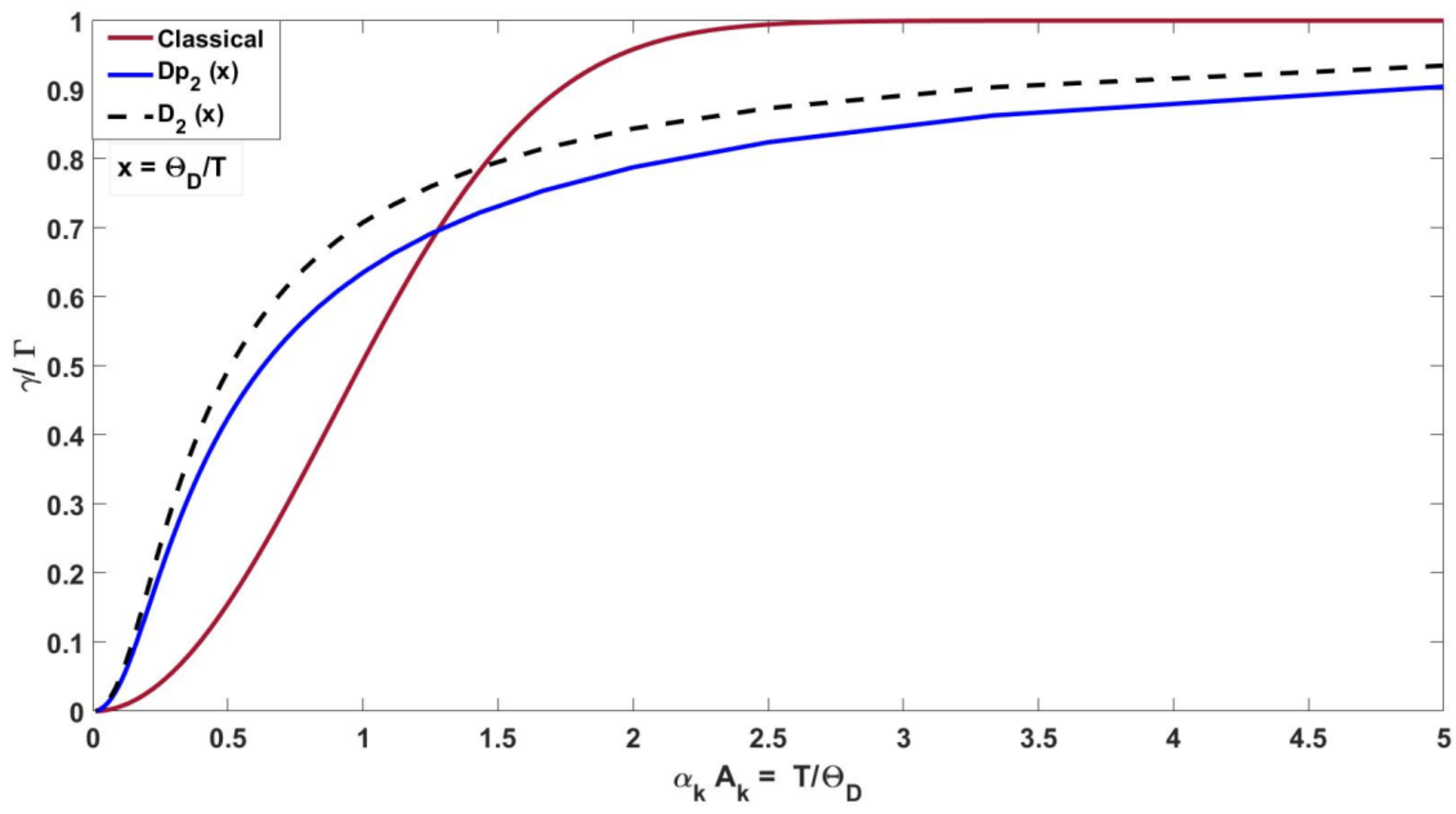

Fig. S1 A comparison between the classical (red) and quantized (blue) kinetic friction coefficients by assuming $(\alpha_k A_k)_{\max} = T / \Theta_D$. The dashed line is the second Debye function, Eq. (46), as the first approximation of the quantized friction coefficient. The blue line is Eq. (49), and the red line is the classical Eq. (21) including unsaturated cases.

Figure S1 shows that assuming $(\alpha_k A_k)_{\max} = T / \Theta_D$, an upper limit around three will saturate the integral of Eq. (22). Moreover, Fig. S1 shows that the classical derivation of the kinetic friction coefficient will reach the asymptotic value of $\Gamma$ faster than the quantized derivation. However, the quantization speeds ups the value of $\gamma$ at lower temperatures up to the Debye temperature and a little further.